\documentclass[12pt,fleqn]{article}
\input{epsfig.sty}
\begin{document}

\begin{center}
{\large \bf Roles of 
quark-pair correlations in baryon structure 
and\\non-leptonic weak transitions of hyperon}

\baselineskip=0.6cm
\vspace {0.7cm}
{\large  K. Suzuki$^a$\footnote{Talk presented at 
1st Asia-Pacific conference on ``Few-body problems in physics'',
Noda/Kashiwa, Japan, August 23-28 (1999)}, 
E. Hiyama$^b$, H. Toki$^a$ and  M. Kamimura$^c$}

\vspace{0.4cm}  
{\em $^b$RCNP, Osaka University, Osaka 567-0047, Japan}\\

\vspace{0.4cm} 
{\em $^a$RIKEN, Saitama 351-0198, Japan}\\

\vspace{0.4cm} 
{\em $^c$Department of Physics, Kyushyu University, Fukuoka 812-8581, 
Japan}

\end{center}

\vspace {0.4cm}

\begin{abstract}
Roles of quark-pair correlations in 
the baryon structure and the hyperon non-leptonic 
weak decay are studied within the non-relativistic constituent 
quark model.  
We construct the SU(3) ground state 
baryons by solving the three body problem rigorously with the 
confinement force and the short range spin-dependent attraction. 
We emphasize the importance of the $s=0$ quark-quark 
correlation to reproduce the 
$\Delta I=1/2$ enhancement of the hyperon decay, and demonstrate 
that resulting static properties as well as the 
decay amplitudes agree with the experiments, if we deal with 
the $s=0$ correlation  properly.  
Special attention is also put on the consequences of the SU(6) 
spin-flavor symmetry breaking due to the $s=0$ correlation.  
Calculated magnetic moments are the almost same as the 
naive SU(6) predictions in spite of the existence of the 
strong correlations.  
\end{abstract}

\section{Introduction}

Properties of light baryons have been extensively studied by 
various models based on the 
constituent quark picture.  
Their results are consistent 
with experiments including applications for the one and two nucleon 
systems,  
although this model involves several adjustable parameters. 
Despite the success of this approach, it is obscure whether or not 
these models correctly describe the quark distributions in 
the baryons, 
and in fact understandings of the non-leptonic weak hyperon 
decay and its $\Delta I = 1/2$ rule are still incomplete.

The $\Delta I = 1/2$ rule implies 
dominance of the $\Delta I =1/2$ transitions on the non-leptonic 
hyperon decay\cite{Review}.  
It is known that relative magnitudes of parity conserving decay 
amplitudes
are reasonably described within the baryon pole approximation,  where  
the weak decay takes place as the two
quark transition process: a $us$-pair in the initial hyperon with 
their total spin $s$ being 0 changes to a $ud$-pair in the 
final state baryon.  
However, if one calculates them using the 
constituent quark models, the absolute value of the 
amplitudes is about a half of the experimental data at 
most\cite{Review,model}.  
We emphasize here, because of the heavy $W$-boson mass, the weak matrix 
elements are 
quite sensitive to the short range quark-quark correlations\cite{Suzuki}.   
Consequently, the failure of the constituent quark model to  
describe the non-leptonic weak transition may indicate the lack of the 
quark correlation in the $s=0$ channel.  

On the other hand, it was pointed out that, from both 
theoretical and phenomenological points of view\cite{diq},  
there exists a strong correlation between quarks in the $s=0$ channel.

In this work we try to clarify the roles of the quark correlation 
in the baryon structure and hyperon non-leptonic weak decays by using 
the constituent quark model.  
We assume the short range spin-dependent 
correlations between quarks together with the confinement force, and 
calculate the baryon masses and other static properties.  
In order to deal with the spin-dependent correlation correctly, 
we must  rigorously solve the three body problem.  
For this purpose, 
we adopt the coupled-rearrangement-channel variational method with 
the Gaussian basis functions which has been developed by 
Hiyama and Kamimura\cite{Hiyama}.

We also focus on the SU(6) breaking effects on baryon properties.  
Introduction of the spin-dependent correlation naturally spoils  
the SU(6) spin-flavor symmetry which is known to work well  
for e.g.~the baryon magnetic moments.  
We shall calculate the magnetic moments to estimate 
the SU(6) breaking effects clearly.

\section{Calculation of weak decay matrix elements}
\label{secstyle}

Let us write the low energy effective weak 
interaction Hamiltonian\cite{Hamiltonian};
\begin{equation}
{\cal H}_{W}={{G_F\sin \theta \cos \theta } \over {\sqrt 2}}
\sum\limits_i^{} {c_i(\mu ^2)\;}O_i + \mbox{h.c.}
\label{whamiltonian}
\end{equation}
where $O_i$ are the quark 4-Fermi operators, and $O_1, O_2$ give dominant 
contributions in our case.  
We take values of $c_i$ given in ref.~\cite{Hamiltonian} at 1GeV$^2$.

Our task here is to evaluate the matrix element
$\langle B_f \; \pi^a | {\cal H}_{W}|  B_i \rangle $ for the 
strangeness changing process 
$B_i \to B_f + \pi^a$.  
The PCAC relation and soft pion theorem are suitable to deal with the strong 
interacting pion-nucleon system.  
Using them, one can find the baryon pole formula to the parity 
conserving amplitudes.   
For example, $\Lambda^0 \to n + \pi ^0$ pole amplitude is given by,
\begin{eqnarray}
{{M_N+M_\Lambda } \over {f_\pi }}\left[ {G_{nn}^{\pi 0}
{1 \over {M_\Lambda -M_N}}\left\langle {n\left|{\cal H}_{W}
  \right|\Lambda } 
\right\rangle 
+\left\langle {n\left| {\cal H}_{W} \right|\Sigma ^0} \right\rangle 
{1 \over {M_N - M_\Sigma }}G_{\Lambda \Sigma }^{\pi 0}} \right]
\label{lambdapole}
\end{eqnarray}
where $\left\langle {n\left|{\cal H}_{W}
  \right|\Lambda } \right\rangle $ and 
$\left\langle {n\left| {\cal H}_{W} \right|\Sigma ^0} \right\rangle $ 
are the matrix elements of eq.~(\ref{whamiltonian}) with appropriate 
baryon states, and $G_{B \; B' }^{\pi a}$ denote the axial vector 
coupling constants which gives probabilities for the 
pion emission $B \to B' + \pi^a$ and are constrained by experiments.  

We come to determine the matrix elements of 
${\cal H}_{W}$, $\langle n | {\cal H}_{W} | \Lambda \rangle$ and 
$\langle p | {\cal H}_{W} | \Sigma^+ \rangle$.  
We recall the quark models such as Harmonic Oscillator 
model or MIT bag model give much smaller values for these 
matrix elements than the data\cite{model}.  
It is instructive to rewrite the $V-A$ operators $O_1, \, O_2$ in the 
non-relativistic limit in the coordinate space as
\begin{equation}
O_1, \, O_2 \to 
a_d^\dag \,  a_u^\dag (1 - \vec \sigma_u \cdot \vec \sigma_s)
\, \,  \delta^{(3)} (\vec r_{us}) \, \, a_u  a_s
\end{equation}
where $a_i$, $ a_i^\dag$ are annihilation and creation operators of 
quarks with 
flavor $i$.  Presence of the spin-projection operator 
$(1 - \vec \sigma_u \cdot \vec \sigma_s)$ tells us that the 
weak transition is generated by the two 
body process between spin-0 quark pairs; $(us)^0 \to (ud)^0$, 
which 
guarantees the $\Delta I=1/2$ dominance 
on the non-leptonic hyperon decays due to the antisymmetrization of 
the quark-pairs.  
Now it is clear that this decay amplitude is  sensitive to the 
correlation of the spin-0 quark pair in the baryons.  
The standard constituent quark model never incorporates such a 
quark-quark correlation properly.  


\begin{figure}[htb]
\begin{minipage}[c]{77mm}
\psfig{file=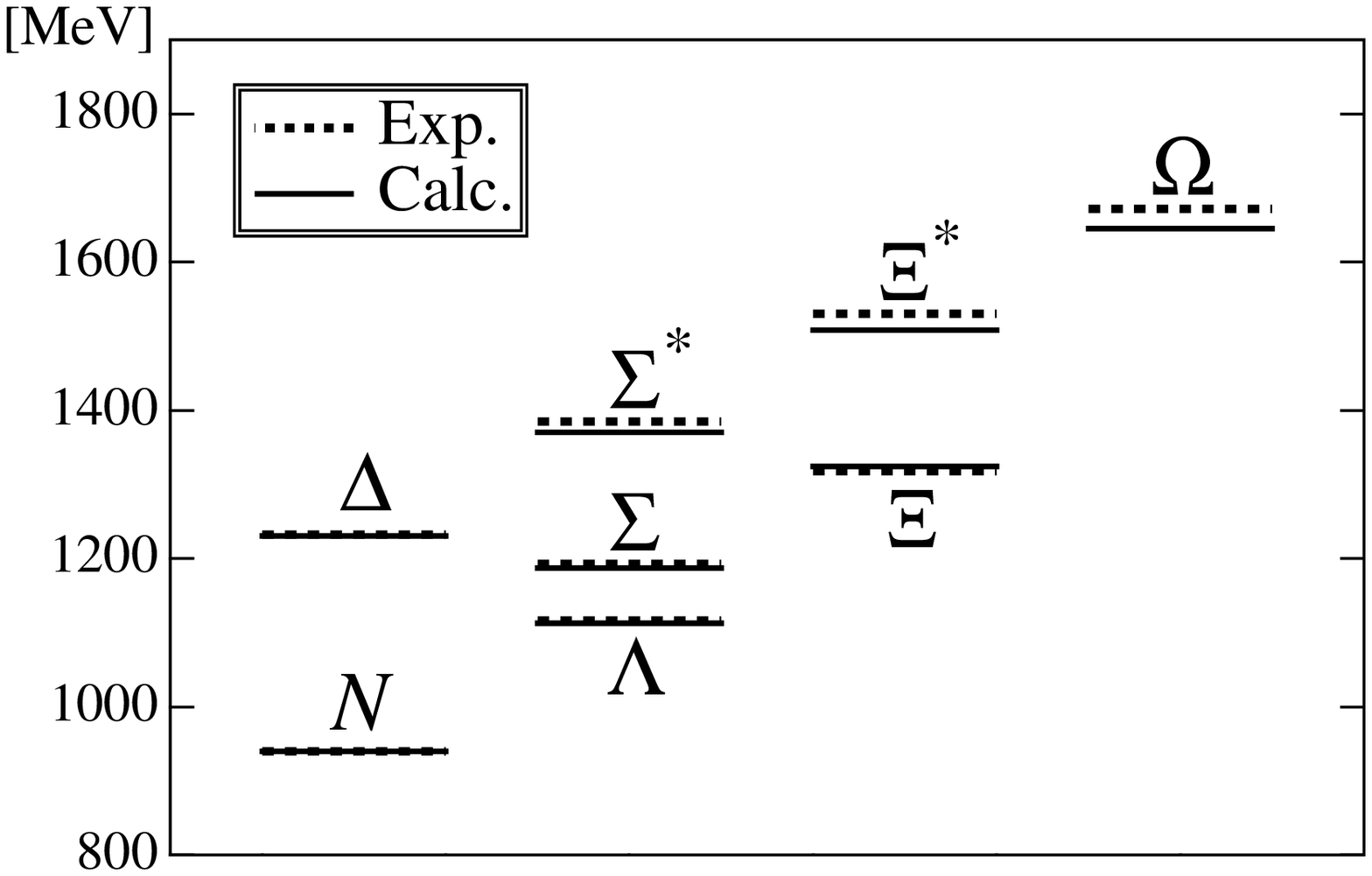,height=1.8in} 
\caption{SU(3) baryon mass spectrum}
\label{fig1}
\end{minipage}
\hspace{\fill}
\begin{minipage}[c]{37mm}
\psfig{file=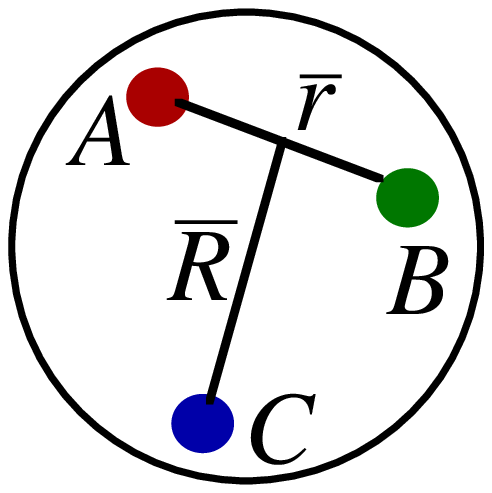,height=1.1in} 
\caption{Configuration of three quarks in nucleon}
\label{fig2}
\end{minipage}
\end{figure}

\vspace{-0.5cm}

\section{Constituent quark model with the spin-dependent correlations}

We phenomenologically introduce the effective 
Hamiltonian which includes the confinement force and the spin-dependent 
part as; 
\begin{eqnarray}
{\cal H} &=&\sum\limits_i {{{p_i^2} \over {2m_i}}}+
\sum\limits_{i<j} {{1 \over 2}K\left( {\vec r_i- \vec 
r_j} \right)^2} +\sum\limits_{i<j} V_{S} (ij) + V_0 \\
V_{S}(ij) &=&\left\{ {\matrix{{0\quad (s=1\;\mbox{pair})}\cr
{\; {{{C_{SS}} \over {m_i m_j}}}
\mbox{Exp}\left[ {-\left( {\vec r_i-\vec 
r_j} \right)^2  / \beta ^2} \right]}\cr
}} \right.\quad (s=0\;\mbox{pair})
\end{eqnarray}
where $m_i$ are the constituent quark masses,  and $K, C_{SS}, \beta$ 
are the model parameters.  
$V_0$ contributes to the over all shift of the resulting spectrum 
and is chosen to adjust the energy of the lowest state to the nucleon mass.   
Constituent quark masses are taken to be 
$m_u = m_d = 330 \mbox{MeV}$ and $m_s = 510 \mbox{MeV}$.

Using this Hamiltonian, we shall solve non-relativistic three 
body problem rigorously.  
We use the coupled-rearrangement-channel 
variational method with Gaussian
basis functions\cite{Hiyama}. 
We assume the isospin symmetry between $u$ and $d$ quarks, and solve the 
three body problem without further approximations or assumptions.  
The strange quark is explicitly distinguished from light $u,d$ quarks.  

We shall fix the model parameters so as to reproduce the 
nucleon and $\Delta$ masses, proton radius and 
the $\Sigma ^+ \to n \pi^+$ amplitude, since this 
decay is completely dominated by the baryon pole diagrams 
(see Table 1).   
We obtain the parameters $K=0.005 \mbox{GeV}^3$, $\beta = 
0.5 \mbox{fm}$ and $C_{SS} / m_u^2 = 1.4 \mbox{GeV}$.

Resulting mass spectrum is shown in Fig.1. 
It could be possible to obtain a better agreement by modifying the potential 
or parameters, but the present results are enough for our purpose.  
The effect of the attractive correlation can be seen clearly by 
introducing the average distances of the 
Jacobi coordinate $\bar r$ and $\bar R$ 
defined in Fig.2, where all 
possible rearrangement channels are transformed to 
the configuration of Fig.2.  
We find, for the nucleon,  $\bar r = 0.92 \mbox{fm}$ and 
$\bar R = 0.97 \mbox{fm}$
when the total spin $s$ of the quark pair $A$ and $B$ is zero, while 
$\bar r = 1.1 \mbox{fm}$ and $\bar R = 0.81 \mbox{fm}$ in the  $s=1$ case.  
Apparently, the quark correlation modifies quark distribution in the nucleon.
The value of the wave function 
at origin $\int d^3 R \; d \Omega_r \Psi (r=0,R)$ 
in the  $s=t=0$ case is about three times 
as large as that of the $s=t=1$ case, 
which provides an huge enhancement for the
$\Delta I = 1/2$ weak decay amplitudes.

The matrix elements of the weak Hamiltonian are calculated in terms of 
our wave functions.  
We find $\langle n | {\cal H}_W | \Lambda \rangle = -0.946 
(\times 10^-2$GeV$^{3}$)
as the full result and 
$-0.310$ when we omit the correlation. Similarly,  
$\langle p | {\cal H}_W | \Sigma ^+ \rangle = 2.86 $ with $V_S$, while 
it becomes $0.762$ without  $V_S$.  
In the absence of the correlation  $V_S$, 
a ratio $\langle p | H_W | \Sigma^+ \rangle / 
\langle  n | H_W | \Lambda \rangle = -2.45$ shows a perfect agreement 
with the SU(6) 
expectation $\sqrt{6} \simeq - 2.4494\cdots$.  
In the realistic case with $V_S$, one can observe the 
substantial enhancement and the SU(6) breaking effect.

With these values, calculated weak transition amplitudes 
are tabulated in Table 1.  
We show the pole contributions only in the second column, and the sum of 
the pole, 
factorization and penguin contributions in the third column to be 
compared with the experiments.  
We find a good agreement for $\Sigma \to N \pi$ decays, while   
the $\Lambda \to N \pi$ amplitude is not enough.  
We note that the small value of the $\Lambda \to N \pi$ 
pole contribution is caused by 
the strong cancellation of the two terms in eq.~(2).  If we vary the 
axial-vector coupling $G$ within the experimental errors ($\sim 10 \%$), 
we find about $40\%$ increase of the $\Lambda \to N \pi$ pole contribution 
with the $\Sigma$ decay amplitudes almost unchanged.

\section{SU(6) symmetry breaking effects on the magnetic moment}

Our wave functions clearly 
violate  the naive SU(6) spin-flavor symmetry for the 
baryon.  In fact, the ratio $\langle p | H_W | \Sigma^+ \rangle / 
\langle  n | H_W | \Lambda \rangle$ becomes $-3.02$, instead of the 
SU(6) value $ -2.45$.  
Thus, we estimate the size of the SU(6) breaking effect 
to be about $20 \%$, which is significant.  
On the other hand, 
it is historically known that the light baryon magnetic moments are well 
reproduced 
in the naive quark model by virtue of the SU(6) spin-flavor symmetry.  
Hence, it is important to examine 
the SU(6) breaking by calculating the magnetic moments.

Our results are shown in Table 2.  
It is manifest that the results are almost unchanged even after 
introducing the spin-dependent correlations.  The differences are of 
order of a few $\%$ in any cases.  
It seems that the global baryon properties such as magnetic moments obtained 
by integrating the wave function over space are insensitve to the 
quark correlation, although the local structure of the 
quark wave function is modified substantially.  


\vspace{0.2cm}

\begin{minipage}[c]{70mm}
\noindent
{\bf Table 1} Parity conserving weak transition amplitude 
(in $10 ^{-7}$ unit)
\vspace{0.1cm}
\noindent
\begin{tabular}{l|cc|c}
 & Pole & total & Exp. \\
\hline
$\Sigma^+_0$ & $ 24.0$   &  26.1   & $26.24$ \\
$\Sigma^+_+$ & $ 43.3$ & 43.3    & $41.83 $ \\
$\Lambda ^0 _0$ & $-3.82$ & $-8.84$ & $ -15.61 $
\end{tabular}
%
\end{minipage}
\hspace{\fill}
\begin{minipage}[c]{70mm}

{\bf Table 2} Magnetic Moments

\vspace{0.1cm}
\begin{tabular}{l|ccc}
 & full & no  $V_{S}$ & Exp. \\
\hline
$\mu_p    $      & $2.75$  &  $2.84$  & 2.79\\
$\mu_n    $      & $-1.78$  &  $-1.90$  & $-1.91$\\
$\mu_\Lambda  $  & $-0.60$  &  $-0.61$  & $-0.61$\\
$\mu_{\Sigma^+} $  & $2.67$  &  $2.73$  & 2.46\\
$\mu_{\Sigma^-}$   & $-1.05$  &  $-1.06$  & $-1.16$\\
$\mu_{\Xi^0}   $   & $-1.40$  &  $-1.45$  & $-1.25$\\
\end{tabular}

\end{minipage}

\vspace{-0.3cm}


\section{Summary}

In conclusion, we have studied the roles of the spin-dependent 
quark correlations 
in the baryon structure.  
We have emphasized that the non-leptonic weak transition of 
the hyperon 
is unique quantity to investigate the quark-quark correlation in the 
spin-0 channel.   
We have solved three body 
problem explicitly using the coupled-rearrangement-channel variational 
method.   
Results for static baryon properties as well as the 
transition amplitudes of the non-leptonic hyperon decay 
reasonably agree with the empirical values.  
We have also discussed the SU(6) breaking effects on the 
baryon properties, 
and pointed out that this symmetry is sill useful for the static baryon 
properties, although local behavior of the quark wave function 
considerably departs from the SU(6) expectation.

\end{document}